\documentclass{article}

\def\xxinput#1{\input#1}

\xxinput{vsolj02.sty}

\usepackage[dvipdfmx]{graphicx}
\usepackage[comma,colon]{natbib}

\usepackage[OT2,T1]{fontenc}

\def\cite{\citealt}
\setcitestyle{aysep={}}

\newcounter{author}
\def\altaffilmark#1{$^{#1}$}
\def\altaffiltext#1{$^{#1}$\,}

\def\authorcount#1#2{{\refstepcounter{author}\label{#1}
                     \altaffiltext{\ref{#1}}{#2}}}

\begin{document}

\begin{center}

\title{Double outbursts in V544 Her and ASASSN-19yt}

\author{
        Taichi~Kato\altaffilmark{\ref{affil:Kyoto}}
}

\authorcount{affil:Kyoto}{
     Department of Astronomy, Kyoto University, Sakyo-ku,
     Kyoto 606-8502, Japan \\
     \textit{tkato@kusastro.kyoto-u.ac.jp}
}

\end{center}

\begin{abstract}
\xxinput{abst.inc}
\end{abstract}

   \citet{kat19csind} reported a superoutburst of the SU UMa star
CS Ind, which showed a long precursor outburst.
The overall behavior looked like a ``double outburst'', which
is sometimes seen in WZ Sge stars (see e.g., \cite{kat15wzsge})
evolved after the period bounce
\citep{kat13j1222,neu17j1222,kim18asassn16dt16hg}.

\begin{figure*}
\begin{center}
\includegraphics[width=16cm]{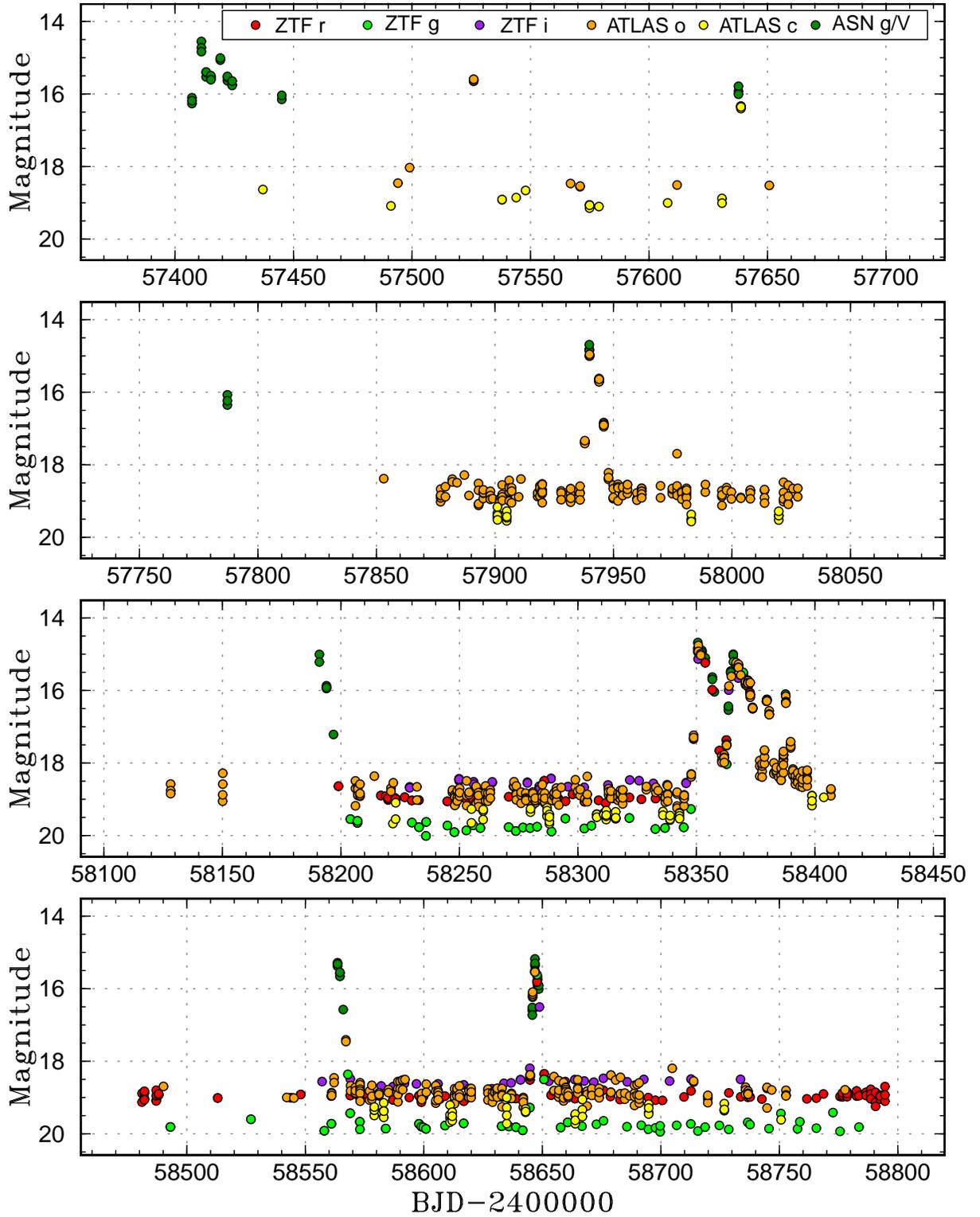}
\caption{
   Light curve of V544 Her in 2016--2020.
ASN refers to ASAS-SN observations.
Only positive detections are shown and all other ASAS-SN
observations were upper limits.
}
\label{fig:lc1}
\end{center}
\end{figure*}

\begin{figure*}
\begin{center}
\includegraphics[width=16cm]{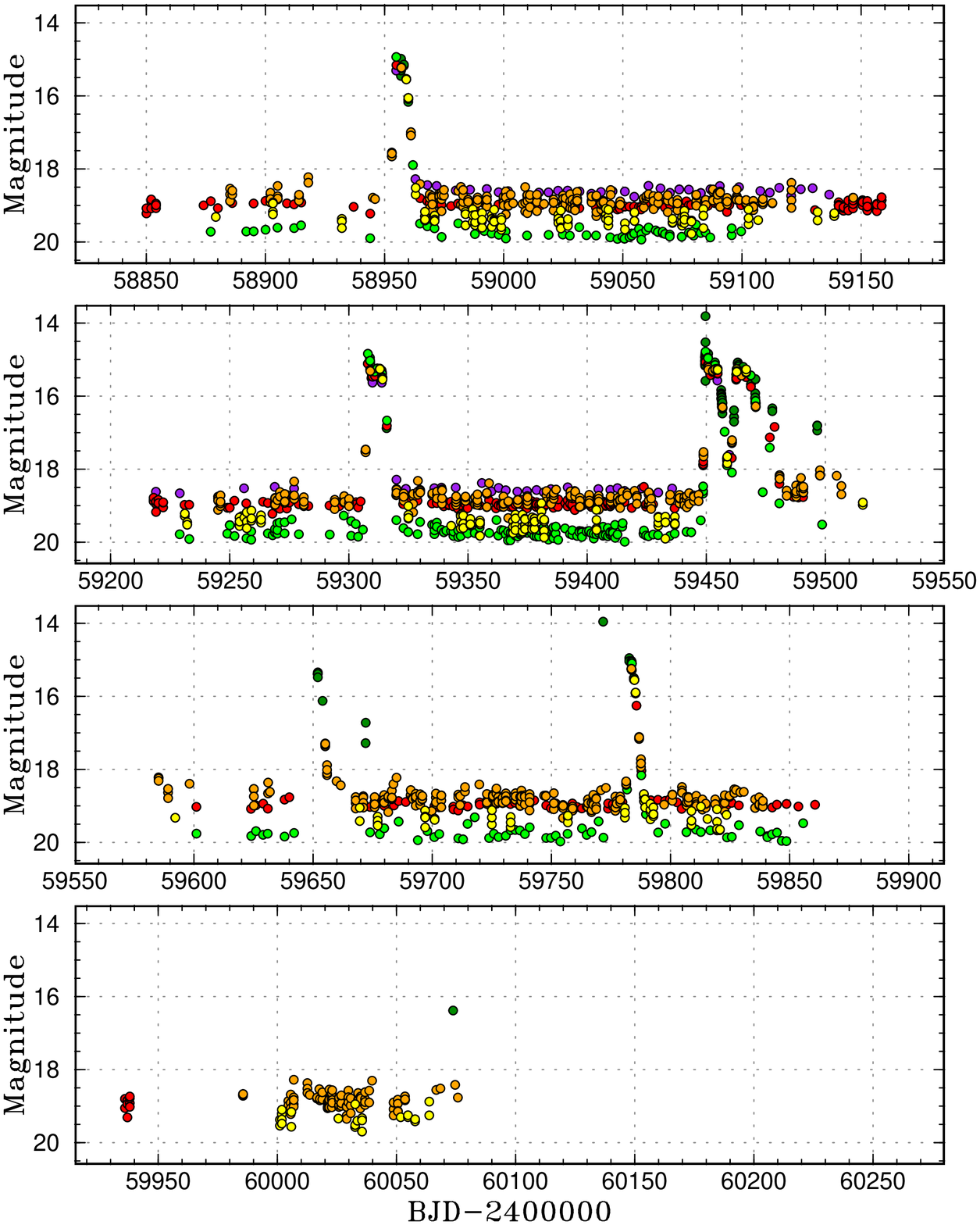}
\caption{
   Light curve of V544 Her in 2020--2023.
The symbols are the same as in figure \ref{fig:lc1}.
}
\label{fig:lc2}
\end{center}
\end{figure*}

   In \citet{kat19csind}, V544 Her was introduced as a dwarf nova
with a double outburst and the similarity with CS Ind
was discussed.  \citet{kat19csind} suggested that this would
be a rare phenomenon based on the absence of a similar
outburst in the past data.
Using Zwicky Transient Facility
(ZTF: \cite{ZTF})\footnote{
   The ZTF data can be obtained from IRSA
$<$https://irsa.ipac.caltech.edu/Missions/ztf.html$>$
using the interface
$<$https://irsa.ipac.caltech.edu/docs/program\_interface/ztf\_api.html$>$
or using a wrapper of the above IRSA API
$<$https://github.com/MickaelRigault/ztfquery$>$.
} and Asteroid Terrestrial-impact
Last Alert System (ATLAS: \cite{ATLAS}) forced photometry
\citep{shi21ALTASforced} observations, I noticed that
V544 Her showed a double outburst again in 2021, contrary
to the expectation in \citet{kat19csind}.
The overall light curve of the past eight seasons
using the ZTF, ATLAS and All-Sky Automated Survey for Supernovae
(ASAS-SN, \cite{ASASSN}) data is shown in figures
\ref{fig:lc1} and \ref{fig:lc2}.  Only positive detections
are shown and all other ASAS-SN observations were
upper limits.  There are several solitary ASAS-SN detections
(such as the recent BJD 2460074 one), but they were likely
noises.

\begin{figure*}
\begin{center}
\includegraphics[width=16cm]{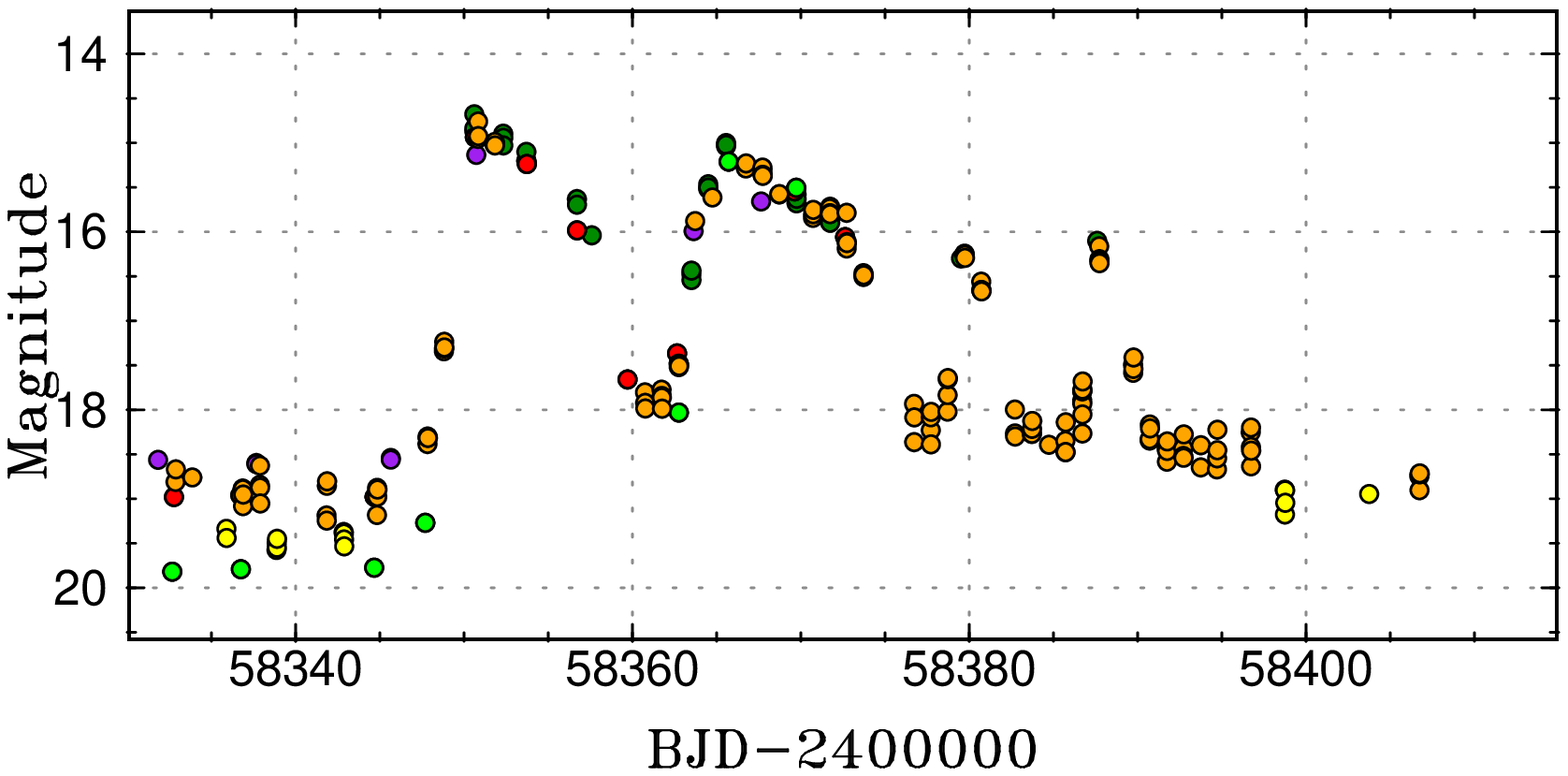}
\caption{
   Double outburst and rebrightenings in V544 Her in 2018.
The symbols are the same as in figure \ref{fig:lc1}.
}
\label{fig:lcout1}
\end{center}
\end{figure*}

\begin{figure*}
\begin{center}
\includegraphics[width=16cm]{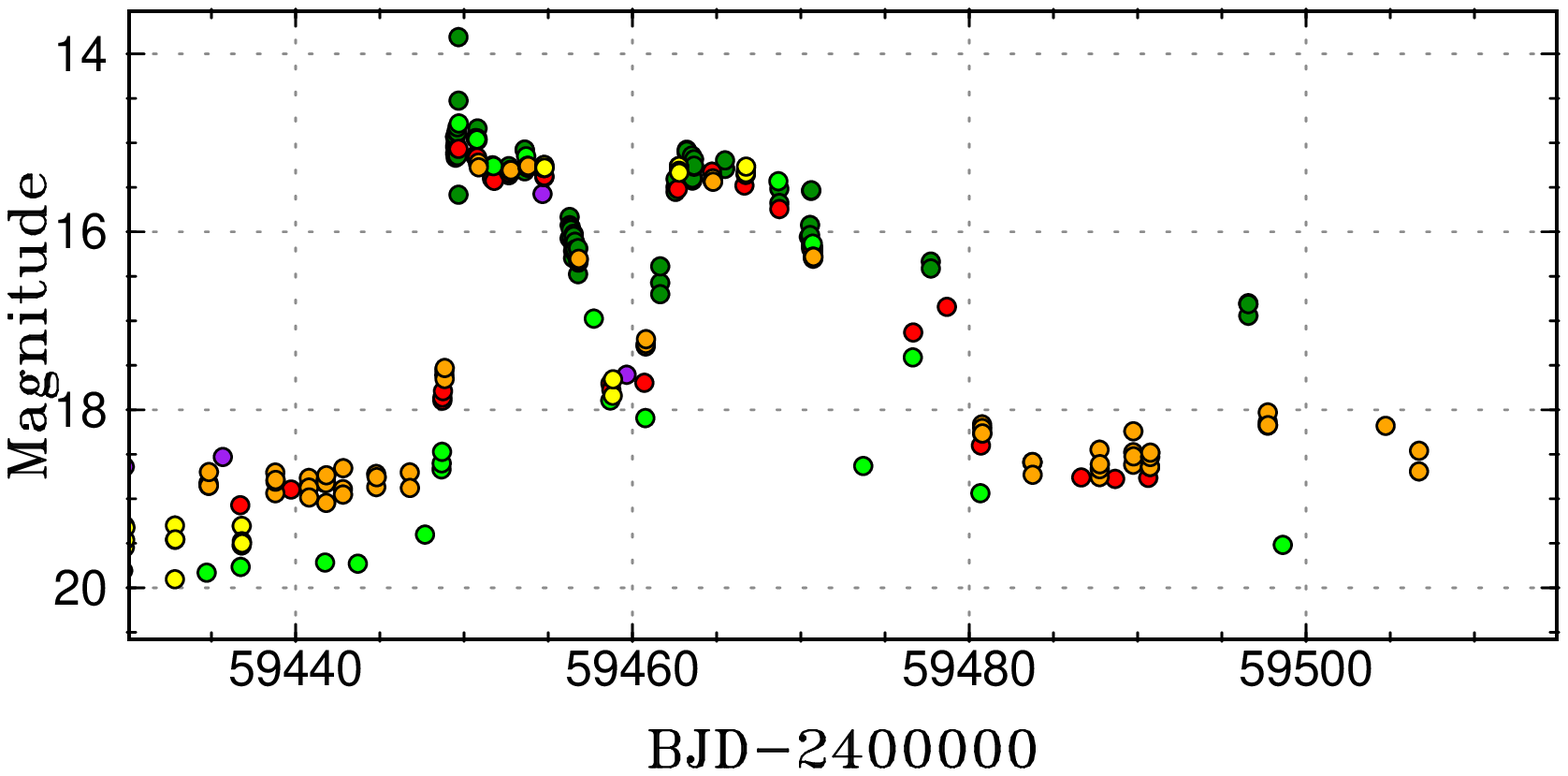}
\caption{
   Double outburst and rebrightenings in V544 Her in 2021.
The symbols are the same as in figure \ref{fig:lc1}.
}
\label{fig:lcout2}
\end{center}
\end{figure*}

   The double outbursts in 2018 and 2021 are shown in detail
in figures \ref{fig:lcout1} and \ref{fig:lcout2}, respectively.
They were very similar: first long outburst, dip,
second long outburst and two short rebrightenings.
These outbursts very much resemble the double outburst
recorded in the SU UMa star CS Ind, as already reported.
The red color of V544~Her [an orbital period over 0.2~d was
inferred from SDSS colors \citep{kat12DNSDSS}; 
$BP$=19.59, $RP$=18.27 \citep{GaiaDR3}]
in quiescence, however, appears to be inconsistent with
an SU UMa star with a short orbital period.
All ZTF and ATLAS observations were snapshtot data and
it was not possible to make a period analysis to detect
possible superhumps.  It looks like, however, that no
large-amplitude (such as 0.2--0.3~mag) superhumps were
present.  Although the 2021 outburst was also detected by
VSOLJ and VSNET \citep{VSNET} members, the detections were
late during the initial long outburst and no special
attention was paid since the object already appeared to be
fading and since the season was too late to
make long time-resolved photometry.

   The first outburst in 2016 in figure \ref{fig:lc1}
might have been a complex one, although the limited quality
of the ASAS-SN data and still limited coverage by ATLAS
made it impossible to examine it in detail.
If double outbursts in this system occur relatively
frequently (such as once in three years), we may have
a more frequent chance to determine the nature of
a double outburst in this system than previously thought.
There have recently been an increasing number of
discoveries of SU UMa stars with long orbital periods
[e.g., SDSS J094002.56$+$274942.0 \citep{kat23j0940};
ASASSN-15cm \citep{kat23asassn15cm} and
BO Cet \citep{kat21bocet,kat23bocet}] and V544 Her
may join this group.  It would be, however, worth noting
that some outbursts in dwarf novae have yet unexplained
dips, such as in MASTER OT J055845.55$+$391533.4
\citep{kat23j0558} and the phenomenon in V544 Her
may not be related to SU UMa-type one.

\begin{figure*}
\begin{center}
\includegraphics[width=16cm]{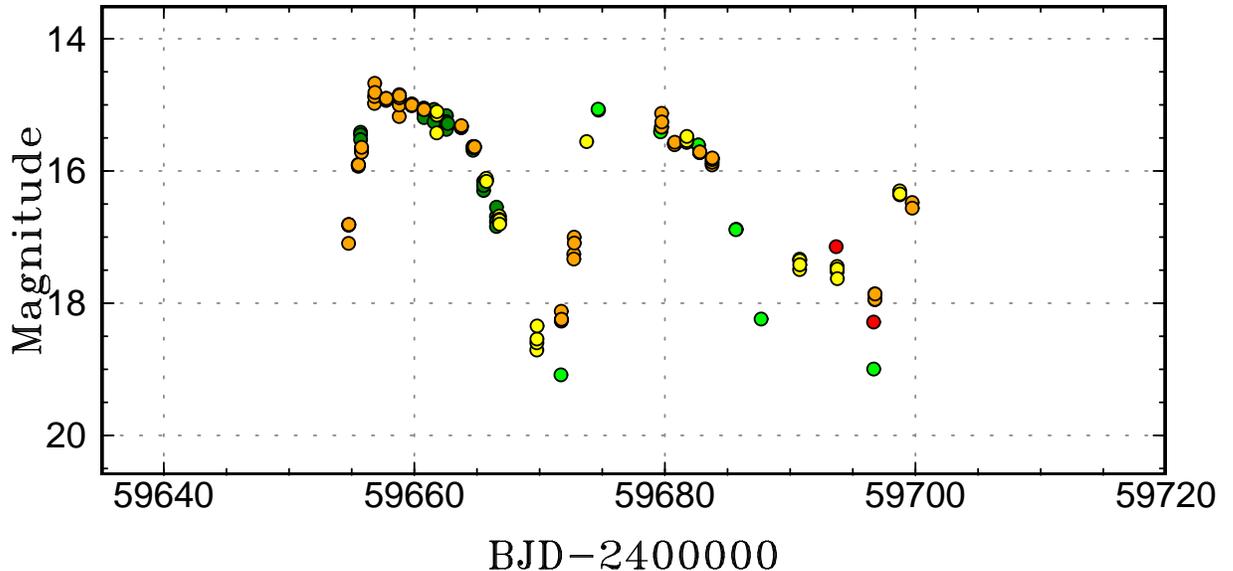}
\caption{
   Double outburst and rebrightenings in ASASSN-19yt in 2022.
This outburst occurred near the end of the season
and ZTF did not cover the quiescence before this outburst.
There were no observations after BJD 2459700.
The symbols are the same as in figure \ref{fig:lc1}.
}
\label{fig:lc19yt1}
\end{center}
\end{figure*}

\begin{figure*}
\begin{center}
\includegraphics[width=16cm]{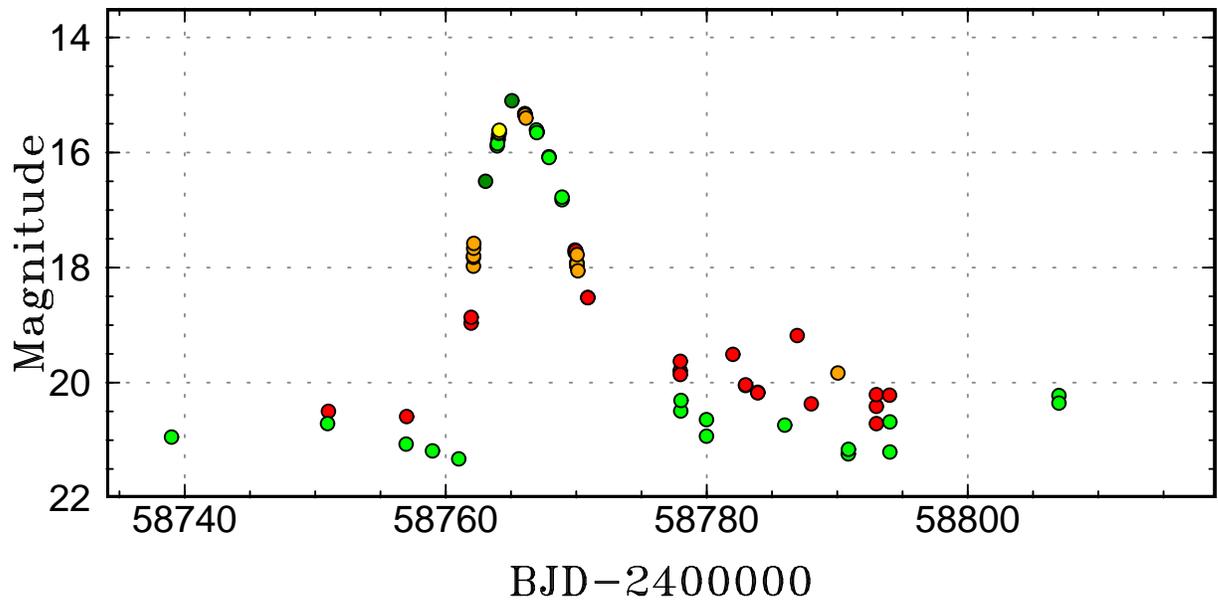}
\caption{
   Long outburst in ASASSN-19yt in 2019.
The symbols are the same as in figure \ref{fig:lc1}.
}
\label{fig:lc19yt2}
\end{center}
\end{figure*}

   I also noticed that ASASSN-19yt showed a double outburst
and rebrightenings in 2022 very similar to V544~Her
(figure \ref{fig:lc19yt1}).  The same object showed
a (relatively) long outburst in 2019 (figure \ref{fig:lc19yt2}).
This outburst rose slowly and the overall symmetric
shape resembled that of an SS Cyg star, although
the maximum fading rate was larger than in SS Cyg stars.
Tonny Vanmunster obtained a single-night run and did not
detect superhumps.
If this object is indeed an SU UMa star, the morphology
of the 2019 outburst would challenge our knowledge
in SU~UMa stars, and, inversely, if this object is
an SS Cyg star, what causes a double outburst
and rebrightenings would become an unsolved problem
in dwarf novae.  Please also remember that there was
also an unusual case of a long outburst and rebrihgtenings
in PY Per \citep{kat22pyper},
whose nature has not yet been clarified.

\section*{Acknowledgements}

This work was supported by JSPS KAKENHI Grant Number 21K03616.
The author is grateful to the ZTF, ATLAS and ASAS-SN teams
for making their data available to the public.
I am also grateful to Naoto Kojiguchi for helping downloading
ZTF data and Tonny Vanmunster for providing observations
of ASASSN-19yt.

Based on observations obtained with the Samuel Oschin 48-inch
Telescope at the Palomar Observatory as part of
the Zwicky Transient Facility project. ZTF is supported by
the National Science Foundation under Grant No. AST-1440341
and a collaboration including Caltech, IPAC, 
the Weizmann Institute for Science, the Oskar Klein Center
at Stockholm University, the University of Maryland,
the University of Washington, Deutsches Elektronen-Synchrotron
and Humboldt University, Los Alamos National Laboratories, 
the TANGO Consortium of Taiwan, the University of 
Wisconsin at Milwaukee, and Lawrence Berkeley National Laboratories.
Operations are conducted by COO, IPAC, and UW.

The ztfquery code was funded by the European Research Council
(ERC) under the European Union's Horizon 2020 research and 
innovation programme (grant agreement n$^{\circ}$759194
-- USNAC, PI: Rigault).

This work has made use of data from the Asteroid Terrestrial-impact
Last Alert System (ATLAS) project.
The ATLAS project is primarily funded to search for
near earth asteroids through NASA grants NN12AR55G, 80NSSC18K0284,
and 80NSSC18K1575; byproducts of the NEO search include images and
catalogs from the survey area. This work was partially funded by
Kepler/K2 grant J1944/80NSSC19K0112 and HST GO-15889, and STFC
grants ST/T000198/1 and ST/S006109/1. The ATLAS science products
have been made possible through the contributions of the University
of Hawaii Institute for Astronomy, the Queen's University Belfast, 
the Space Telescope Science Institute, the South African Astronomical
Observatory, and The Millennium Institute of Astrophysics (MAS), Chile.

\section*{List of objects in this paper}
\xxinput{objlist.inc}

\section*{References}

We provide two forms of the references section (for ADS
and as published) so that the references can be easily
incorporated into ADS.

\newcommand{\noop}[1]{}\newcommand{\hyphalt}{-}

\renewcommand\refname{\textbf{References (for ADS)}}

\xxinput{v544heraph.bbl}

\renewcommand\refname{\textbf{References (as published)}}

\xxinput{v544her.bbl.vsolj}

\end{document}